\documentclass[twocolumn,showpacs,preprintnumbers,amsmath,amssymb]{revtex4}
\usepackage{graphicx}% Include figure files
\usepackage{dcolumn}% Align table columns on decimal point
\usepackage{bm}% bold math

\begin{document}

\title{$\alpha$ decay half-lives of new superheavy nuclei within a generalized liquid drop model}
\author{Hongfei Zhang$^{1,2,3}$}
\email{zhanghongfei@impcas.ac.cn}
\author{Wei Zuo$^{1,2}$}
\author{Junqing Li$^{1,2}$}
\author{G. Royer$^{4}$}

\affiliation{\footnotesize $^1$Institute of Modern Physics,
Chinese Academy of Science, Lanzhou 730000, China}
\affiliation{\footnotesize $^2$Research Center of Nuclear Theory
of Laboratory of Heavy Ion Accelerator of Lanzhou, Lanzhou 730000,
China} \affiliation{\footnotesize$^3$Graduate school of Chinese
Academy of Sciences, Beijing 100039, China}
\affiliation{\footnotesize $^4$Laboratoire Subatech, UMR:
IN2P3/CNRS-Universit\'e-Ecole des Mines 4 rue A. Kastler,
 44307 Nantes Cedex 03, France}

\date{\today}

\begin{abstract}
The $\alpha$ decay half-lives of the recently produced isotopes of
the 112, 114, 116 and $118$ nuclei and decay products have been calculated
in the quasi-molecular shape path using the experimental $Q_{\alpha}$ value
and a Generalized Liquid Drop Model
including the proximity effects between nucleons in the neck or the gap between
the nascent fragments.
  Reasonable estimates are obtained for the observed
$\alpha$ decay  half-lives. The results are compared with
calculations using the Density-Dependent M3Y effective interaction
and the Viola-Seaborg-Sobiczewski formulae.
 Generalized Liquid Drop Model predictions are
provided for the $\alpha$ decay half-lives of other superheavy
nuclei using the Finite Range Droplet Model $Q_{\alpha}$ and
compared with the values derived from the VSS formulae.

\end{abstract}

\maketitle

The synthesis of superheavy elements has advanced using both cold \cite{Hof00} and warm
fusion reactions.
Recently \cite{Og99,Og03,Og05}, isotopes of the elements 112, 114, 116 and 118
 have been produced in fusion-evaporation reactions
at low excitation energies by irradiations of the $^{233,238}$U,
$^{242}$Pu, $^{248}$Cm and $^{249}$Cf targets with $^{48}$Ca
beams. The main decay mode is the $\alpha$ emission and the
$\alpha$ decay energies and half-lives of fourteen new $\alpha$
decaying nuclei have been measured. Some questions have been
raised \cite{Ar00} about these superheavy element findings. In
similar sophisticated experiments at other places \cite{Lo02,Gr05}
the $\alpha$ cascades were not observed.

The pure Coulomb barrier sharply peaked at the touching point does
not allow to determine correctly the partial $\alpha$ decay
half-lives. It is probable that the $\alpha$ decay takes place in
the quasi-molecular shape path where the nucleon-nucleon forces
act strongly during the formation of the neck between the nascent
fragments and after the separation and a proximity energy term
must be added in the usual development of the liquid-drop model
\cite{Bl77}. The generalized liquid drop model (GLDM) which
includes such a proximity energy term has allowed to describe the
fusion \cite{Roy85} , fission \cite{royzb02}, light nucleus
\cite{rm01} and $\alpha$ emission \cite{Roy00,Roy04} processes.

The purpose of this work is to determine the partial $\alpha$
decay half-lives of these superheavy elements within this GLDM
from the experimental $Q_{\alpha}$ values using the WKB
approximation and to compare with the experimental data and the
calculations with the Density-Dependent M3Y (DDM3Y) effective
interaction \cite{Ch06} and the Viola-Seaborg formulae with
Sobiczewski constants (VSS) \cite{So89}. Finally predictions
within the GLDM and VSS formulae are given for the partial
$\alpha$ decay half-lives of the still non observed superheavy
nuclei ranging from Sg to $Z=120$.

For a deformed nucleus, the macroscopic GLDM energy is defined as
\cite{Roy85}.
\begin{equation}\label{etot}
E=E_{V}+E_{S}+E_{C}+E_{\text{Rot}}+E_{\text{Prox}}.
\end{equation}
When the nuclei are separated:
\begin{equation}\label{ev}
E_{V}=-15.494\left \lbrack (1-1.8I_1^2)A_1+(1-1.8I_2^2)A_2\right
\rbrack \ \textrm{MeV},
\end{equation}
\begin{equation}\label{es}
E_{S}=17.9439\left
\lbrack(1-2.6I_1^2)A_1^{2/3}+(1-2.6I_2^2)A_2^{2/3} \right \rbrack
\ \textrm{MeV},
\end{equation}
\begin{equation}\label{ec}
E_{C}=0.6e^2Z_1^2/R_1+0.6e^2Z_2^2/R_2+e^2Z_1Z_2/r,
\end{equation}
where $A_i$, $Z_i$, $R_i$ and $I_i$ are the mass number, charge
number, radii and relative neutron excesses of the two nuclei. $r$
is the distance between the mass centres. The radii $R_{i}$ are
given by \cite{Bl77}:
\begin{equation}\label{radii}
R_i=(1.28A_i^{1/3}-0.76+0.8A_i^{-1/3}) \ \textrm{fm}.
\end{equation}
This formula allows to follow the experimentally observed increase
of the ratio r$_i$=R$_i$/A$_i^{1/3}$ with the mass; for example,
$r_0=1.13$~fm for $^{48}$Ca and $r_0=1.18$~fm for $^{248}$Cm.

For one-body shapes, the surface and Coulomb energies are
defined as:
\begin{equation}\label{esone}
E_{S}=17.9439(1-2.6I^2)A^{2/3}(S/4\pi R_0^2) \ \textrm{MeV},
\end{equation}
\begin{equation}\label{econe}
E_{C}=0.6e^2(Z^2/R_0) \times 0.5\int
(V(\theta)/V_0)(R(\theta)/R_0)^3 \sin \theta d \theta.
\end{equation}
$S$ is the surface of the one-body deformed nucleus. $V(\theta )$
is the electrostatic potential at the surface and $V_0$ the
surface potential of the sphere.

The rotational energy is determined within the rigid-body ansatz:
%\begin{equation}\label{erot}
    $E_{\text{Rot}}=\frac{\hbar^{2}l(l+1)}{2I_{\bot}}.$
%\end{equation}
The surface energy results from the effects of the surface tension
forces in a half space. When there are nucleons in regard in a
neck or a gap between separated fragments an
additional term called proximity energy must be added to take into
account the effects of the nuclear forces between the close
surfaces. This term is essential to describe smoothly the one-body
to two-body transition and to obtain reasonable fusion barrier
heights. It moves the barrier top to an external position and
strongly decreases the pure Coulomb barrier.
\begin{equation}
E_{\text{Prox}}(r)=2\gamma \int _{h_{\text{min}}} ^{h_{\text{max}}} \Phi \left \lbrack
D(r,h)/b\right \rbrack 2 \pi hdh,
\end{equation}
where $h$ is the distance varying from the neck radius or zero to
the height of the neck border. $D$ is the distance between the
surfaces in regard and $b=0.99$~fm the surface width. $\Phi$ is the
proximity function of Feldmeier \cite{fel79}. The surface
parameter $\gamma$ is the geometric mean between the surface
parameters of the two nuclei or fragments. The combination of the
GLDM and of a quasi-molecular shape sequence has allowed to
reproduce the fusion barrier heights and radii, the fission and
the $\alpha$ and cluster radioactivity data.

For the $\alpha$ emission this very accurate formula simulates the proximity energy \cite{mou01}:
\begin{eqnarray}\label{prox2}
    E_{\text{prox}}(r)=(4\pi\gamma)e^{-1.38(r-R_{\alpha}-R_{d})}[0.6584A^{2/3} \\
-(\frac{0.172}{A^{1/3}}+0.4692A^{1/3})r \nonumber\\
    -0.02548A^{1/3}r^{2}+0.01762r^{3}]\nonumber.
\end{eqnarray}
To obtain the $\alpha$ decay barrier from the contact
point between the nascent $\alpha$ particle and daughter nucleus
it is sufficient to add this proximity energy to the
Coulomb repulsion.

The half-life of a parent nucleus decaying via $\alpha$ emission
is calculated using the WKB barrier penetration probability. In a
unified fission model, the decay constant of the $\alpha$ emitter
is simply defined as
%\begin{equation}\label{constant}
    $\lambda=\nu_{0}P$.
%\end{equation}
The assault frequency $\nu_{0}$ has been taken as
%\begin{equation}\label{mu}
    $\nu_{0}=10^{20}s^{-1}$.
%\end{equation}
The barrier penetrability $P$ is calculated within the action
integral
\begin{equation}\label{penetrability}
    P=exp[-\frac{2}{\hbar}\int_{R_{\text{in}}}^{R_{\text{out}}}\sqrt{2B(r)(E(r)-E(sphere))}].
\end{equation}
The deformation energy (relative to sphere) is small until the rupture point between the fragments
\cite{Roy00} and the two following approximations may
been used:$R_{\text{in}}=R_{d}+R_{\alpha}$ and  $B(r)=\mu$
%\begin{equation}\label{points}
%    R_{\text{in}}=R_{d}+R_{\alpha} ,
%\end{equation}
%and
%\begin{equation}\label{points}
% B(r)=\mu .
%\end{equation}
where $\mu$ is the reduced mass. $R_{out}$ is simply
e$^{2}$Z$_{d}$Z$_{\alpha}$/Q$_{\alpha}$. The partial half-life is
related to the decay constant $\lambda$ by
%M\begin{equation}\label{life}
    $T_{1/2}=\frac{\textrm{ln}2}{\lambda}$ .
%\end{equation}
The $\alpha$ decay half-lives  of the recently produced superheavy nuclei calculated with the GLDM
 using the experimental $Q_{\alpha}$
value and without considering the
rotational contribution are presented in Table 1. The
 results agree reasonably with the experimental data indicating
that a GLDM taking account the proximity effects, the mass
asymmetry, and an accurate nuclear radius is sufficient to
reproduce the $\alpha$ decay potential barriers when the
experimental $Q_{\alpha}$ value is known. The results obtained
with the DDM3Y interaction agree with the experimental data as the
GLDM predictions and largely better than the VSS calculations.
This shows that a double folding potential obtained using M3Y
\cite{Be77} effective interaction supplemented by a zero-range
potential for the single-nucleon exchange is very appropriate
because its microscopic nature includes many nuclear features, in
particular a potential energy surface is inherently embedded in
this description. This double agreement shows that the
experimental data themselves seem to be consistent. For most
nuclei the predictions of the VSS model largely overestimate the
half lives. The blocking effect is probably treated too roughly.

The half live of $^{294}118$ is slightly underestimated in the
three theoretical calculations possibly due to the neutron
submagic number $N=176$. In Ref. \cite{zh04}, it is also pointed
out that for oblate deformed chain of Z$=112$, the shell closure
appears at N$=176$.

Most of the theoretical half lives using GLDM are slightly smaller
than the experimental data. A reason is perhaps that the rotation
of the nuclei is neglected in the present calculations. The term
$\hbar^2 l(l+1)/(2$I$_{\bot})$ in Eq.(\ref{erot}) represents an
additional centrifugal contribution to the barrier which reduces
the tunnelling probability and increases the half lives. A second
reason is that the shell effects and pairing correlation are not
explicitly included in the alpha decay barrier, in spite of their
global inclusion in the decay energy $Q$.

\begin{table*}
\label{table1}
\caption{Comparison between experimental $\alpha$ decay half-lives
\cite{Og05} and results obtained
 with the GLDM, the DDM3Y effective
interaction \cite{Ch06} and the VSS formulae. }
\begin{ruledtabular}
\begin{tabular}{ccccccccc}
Parent & Nuclei & Expt.& Expt.& DDM3Y &GLDM &VSS   \\
$Z$&$A$&$Q$(MeV)&$T_{1/2}$&$T_{1/2}$&$T_{1/2}$&$T_{1/2}$\\ \hline
%&&&&&&& \\
  118 &294 &$11.81\pm0.06$& $1.8^{+75}_{-1.3}$ ms &$0.66^{+0.23}_{-0.18}$ ms &$0.15^{+0.05}_{-0.04}$ ms &$0.64^{+0.24}_{-0.18}$ ms    \\
%&&&&&&& \\
  116 &293  &$10.67\pm0.06$& $53^{+62}_{-19}$ ms  &$206^{+90}_{-61}$ ms &$22.81^{+10.22}_{-7.06}$ ms   &$1258^{+557}_{-384}$ ms     \\
%&&&&&&& \\
  116 &292  &$10.80\pm0.07$& $18^{+16}_{-6}$ ms   &$39^{+20}_{-13}$ ms  &$10.45^{+5.65}_{-3.45}$ ms   &$49^{+26}_{-16}$ ms    \\
%&&&&&&& \\
  116&291 &$10.89\pm0.07$&$6.3^{+11.6}_{-2.5}$ ms  &$60.4^{+30.2}_{-20.1}$ ms &$6.35^{+3.15}_{-2.08}$ ms &$336.4^{+173.1}_{-113.4}$ ms  \\
%&&&&&&& \\
  116 &290   &$11.00\pm0.08$& $15^{+26}_{-6}$ ms  &$13.4^{+7.7}_{-5.2}$ ms &$3.47^{+1.99}_{-1.26}$ ms  &$15.2^{+9.0}_{-5.6}$ ms  \\
%&&&&&&& \\
  114 &289  & $9.96\pm0.06$& $2.7^{+1.4}_{-0.7}$ s  &$3.8^{+1.8}_{-1.2}$ s  &$0.52^{+0.25}_{-0.17}$ s &$26.7^{+13.1}_{-8.7}$ s    \\
%&&&&&&& \\
  114 &288  &$10.09\pm0.07$& $0.8^{+0.32}_{-0.18}$ s  &$0.67^{+0.37}_{-0.27}$ s &$0.22^{+0.12}_{-0.08}$ s &$0.98^{+0.56}_{-0.40}$ s   \\
%&&&&&&& \\
  114 &287 &$10.16\pm0.06$& $0.51^{+0.18}_{-0.10}$ s &$1.13^{+0.52}_{-0.40}$ s  &$0.16^{+0.08}_{-0.05}$ s &$7.24^{+3.43}_{-2.61}$ s  \\
%&&&&&&& \\
  114 &286  &$10.35\pm0.06$& $0.16^{+0.07}_{-0.03}$ s &$0.14^{+0.06}_{-0.04}$ s &$0.05^{+0.02}_{-0.02}$ s &$0.19^{+0.08}_{-0.06}$ s \\
%&&&&&&& \\
  112 &285  &$9.29\pm0.06$& $34^{+17}_{-9}$ s &$75^{+41}_{-26}$ s &$13.22^{+7.25}_{-4.64}$ s &$592^{+323}_{-207}$ s   \\
%&&&&&&& \\
  112 &283 & $9.67\pm0.06$& $4.0^{+1.3}_{-0.7}$ s  &$5.9^{+2.9}_{-2.0}$ s &$0.95^{+0.48}_{-0.32}$ s &$41.3^{+20.9}_{-13.8}$ s   \\
%&&&&&&& \\
  110 &279  &$9.84\pm0.06$& $0.18^{+0.05}_{-0.03}$ s &$0.40^{+0.18}_{-0.13}$ s &$0.08^{+0.04}_{-0.02}$ s &$2.92^{+1.4}_{-0.94}$ s  \\
%&&&&&&& \\
  108 &275  &$9.44\pm0.07$& $0.15^{+0.27}_{-0.06}$ s &$1.09^{+0.73}_{-0.40}$ s &$0.27^{+0.16}_{-0.10}$ s &$8.98^{+5.49}_{-3.38}$ s  \\
%&&&&&&& \\
  106 &271 &$8.65\pm0.08$& $2.4^{+4.3}_{-1.0}$ min &$1.0^{+0.8}_{-0.5}$ min &$0.33^{+0.28}_{-0.16}$ min &$8.6^{+7.3}_{-3.9}$ min   \\

\end{tabular}
\end{ruledtabular}
\end{table*}

The experimental $\alpha$ decay half-lives are between the
close theoretical values given by the GLDM and the ones derived from the
VSS formulae. Thus predictions of the $\alpha$ decay half lives with the GLDM and VSS formulae are
possible. In the next calculations the experiential $Q_{\alpha}$ values are taken
from the FRDM \cite{Mo97} which reproduces all known experimental
data of ground state properties of a large number of nuclei and
gives good predictions for nuclei far from the $\beta$ stability
line and the superheavy nucleus region. In Ref. \cite{Roy04} $T_{1/2}$ obtained using
the $Q_{\alpha}$ values given by the Thomas-Fermi model can be found.

\begin{figure*}[t]
\centering
\includegraphics[scale=0.5]{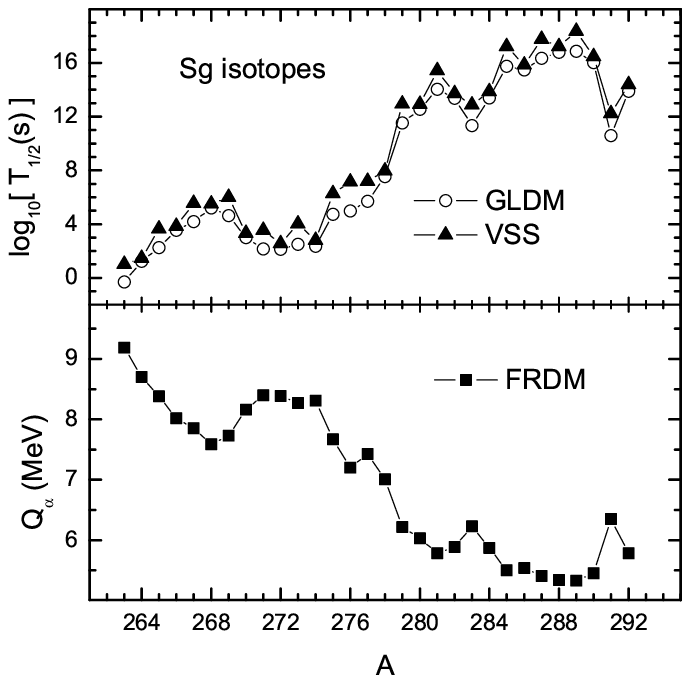}
\includegraphics[scale=0.5]{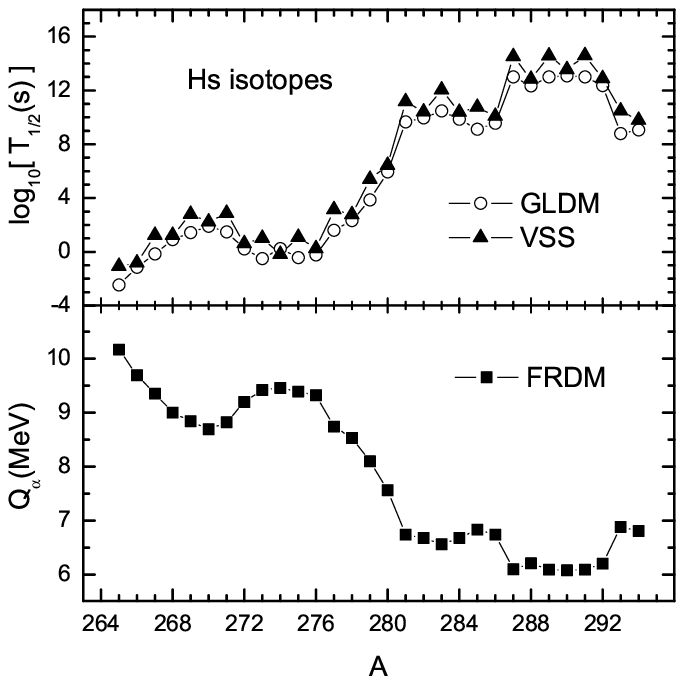}
\includegraphics[scale=0.5]{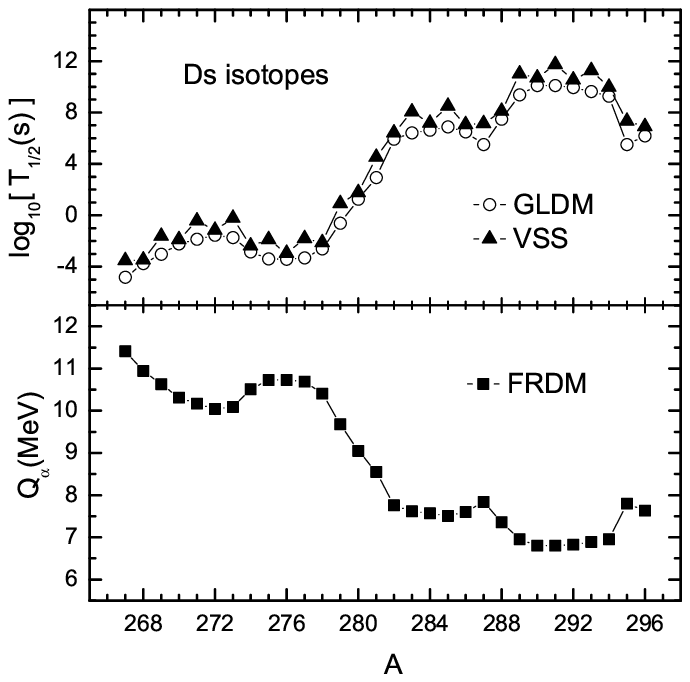}
\includegraphics[scale=0.5]{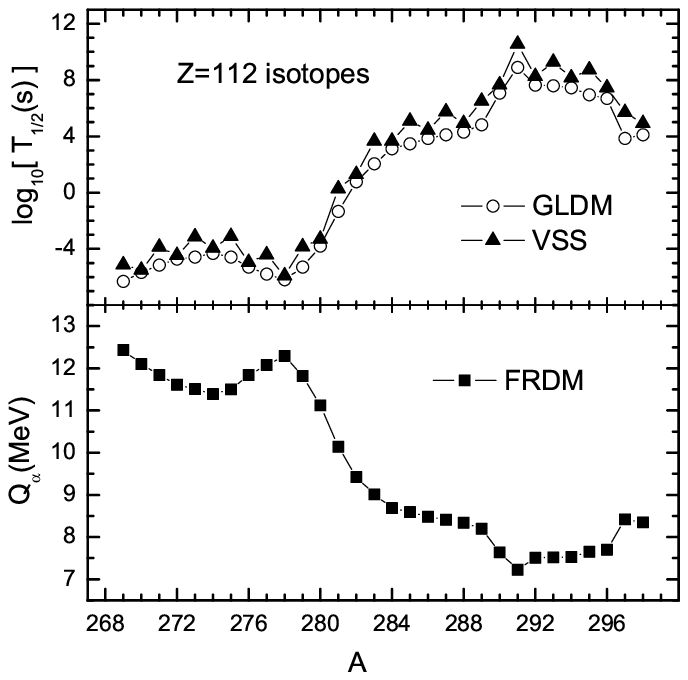}\nonumber\\
\includegraphics[scale=0.5]{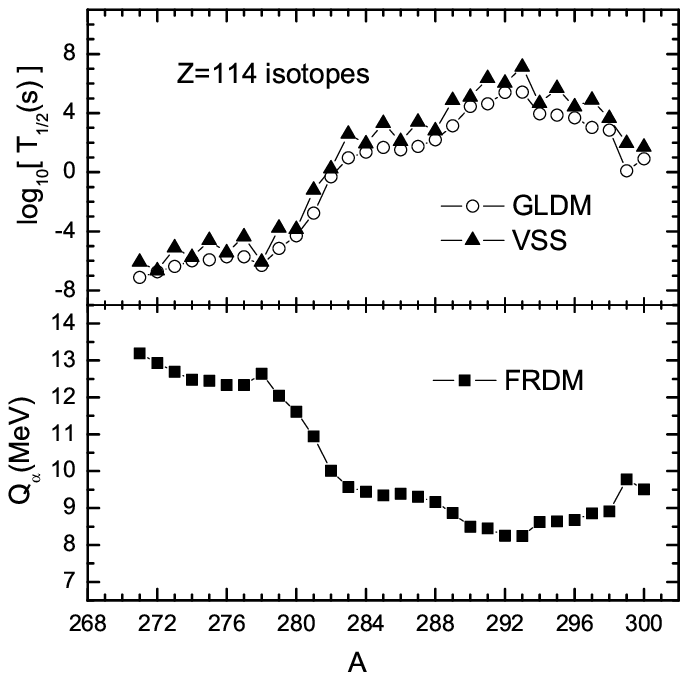}
\includegraphics[scale=0.5]{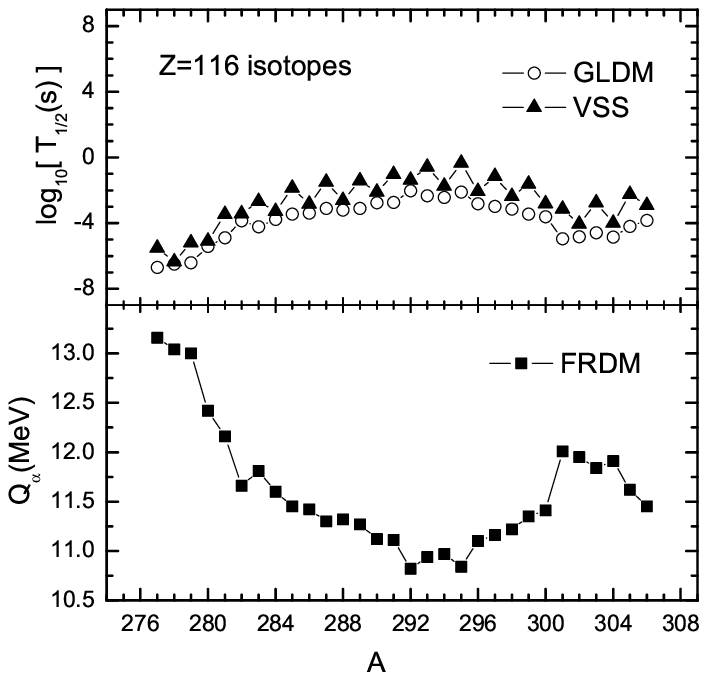}
\includegraphics[scale=0.5]{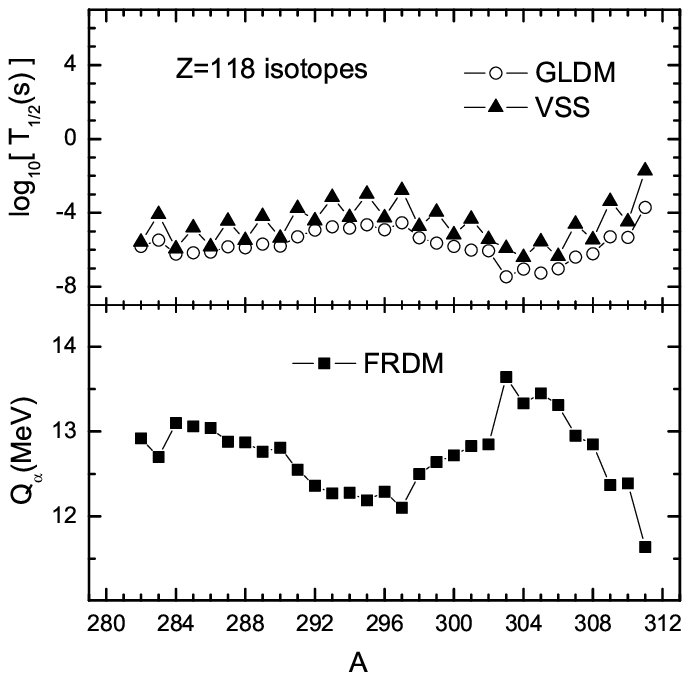}
\includegraphics[scale=0.57]{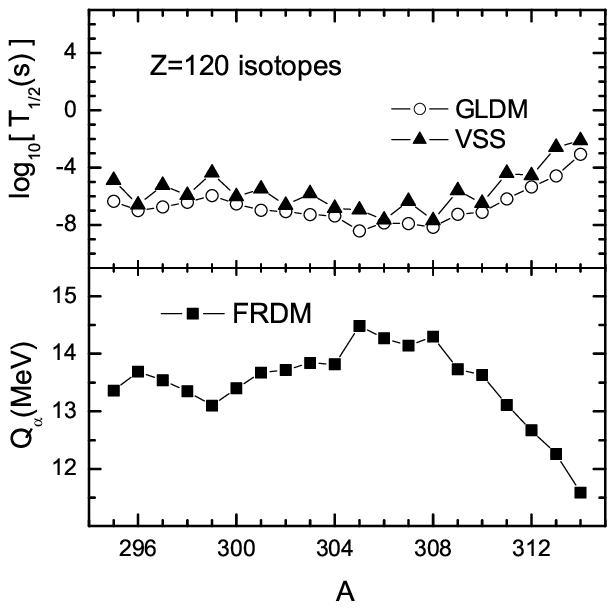}
\caption{Comparison between calculated $\alpha$-decay
     half-lives of Sg, Hs, Ds, 112, 114, 116, 118 and 120 isotopes using the GLDM and the VSS formulae.}
  \label{Fig1}

\end{figure*}

$\alpha$ decay half-lives for $Z=106$ to $Z=120$ isotopes are
shown in Fig.\ref{Fig1}, the open dots indicating the results of
GLDM and the black triangles the ones derived from the VSS
formulae.

The half-life values vary from years to microseconds. A narrow
window exists between the two predictions for each isotope and the
unknown $\alpha$ decay half-lives of SHN may lie in this window
assuming that the FRDM $Q_{\alpha}$ is correct.

The FRDM $Q_{\alpha}$ value explicitly displays a minimum at the
submagic number N$=162$. It is progressively eroded by the neutron
deficiency and the subshell disappears completely from Z$=115$.
This induces a small first hump in the predicted log$_{10}\lbrack
T_{1/2}(s)\rbrack$ curves till Z$=114$. Before this first submagic
number the half-lives increases rapidly with A from Db to Z$=114$
isotopes.
 For the results of VSS formulae, it seems that the turning point is
delayed a little and appears at neutron number N$=163$, for the
blocking effect has been magnified and the half lives are
overestimated. N$=184$ is always a closure shell. Subclosure
or closure shells exist also around N$=176$ for Z$=105$ to
Z$=112$. The most stable nuclei should stand about N$=184$ when the proton
number is higher than Z$=115$, but stand around N$=176$ for lower
Z values.

As a conclusion the half-lives for $\alpha$-radioactivity have
been analyzed in the quasimolecular shape path within a
Generalized Liquid Drop Model including the proximity effects
between nucleons and the mass and charge asymmetry. The results
are in reasonable agreement with the published experimental data
for the alpha decay half-lives of isotopes of charge 112, 114, 116
and 118 and close to the ones derived from the DDM3Y effective
interaction. The experimental $\alpha$ decay half lives stand
between the GLDM calculations and VSS formulae results and the
$\alpha$ decay half-lives of still non-observed superheavy nuclei
have been predicted within the GLDM and VSS approaches and
Q$_{\alpha}$ derived from the FRDM.

This work was supported by the Major State Basic Research
Development Program in China Under Contract No. G2000077400, the
Natural Science Foundation of China (NSFC) under Grant No.
10505016, 10235020, 10235030, 10575119, Knowledge Innovation
Project of the Chinese Academy of Sciences under Grant No
KJCX2-SW-N17£¬and the financial support from DFG of Germany.

%\begin{references}

\end{document}